\documentclass[nofootinbib,twocolumn,preprintnumbers]{revtex4-1}
\pdfoutput=1
\usepackage{amsmath,amsthm,amssymb,multirow,psfrag}
\usepackage{epsfig}
\usepackage{color}
\usepackage{slashed}
\usepackage{soul}
\usepackage{appendix}
\graphicspath{{./Figures/}}

\begin{document}

\def\lsim{\mathrel{\rlap{\lower4pt\hbox{\hskip1pt$\sim$}}
  \raise1pt\hbox{$<$}}}
\def\gsim{\mathrel{\rlap{\lower4pt\hbox{\hskip1pt$\sim$}}
  \raise1pt\hbox{$>$}}}
\newcommand{\vev}[1]{ \left\langle {#1} \right\rangle }
\newcommand{\bra}[1]{ \langle {#1} | }
\newcommand{\ket}[1]{ | {#1} \rangle }
\newcommand{\ev}{ {\rm eV} }
\newcommand{\kev}{{\rm keV}}
\newcommand{\mev}{{\rm MeV}}
\newcommand{\gev}{{\mathrm GeV}}
\newcommand{\tev}{{\rm TeV}}
\newcommand{\mpl}{$M_{Pl}$}
\newcommand{\mw}{$M_{W}$}
\newcommand{\Ft}{F_{T}}
\newcommand{\Zparity}{\mathbb{Z}_2}
\newcommand{\BLambda}{\boldsymbol{\lambda}}
\newcommand{\met}{\;\not\!\!\!{E}_T}
\newcommand{\beq}{\begin{equation}}
\newcommand{\eeq}{\end{equation}}
\newcommand{\bea}{\begin{eqnarray}}
\newcommand{\eea}{\end{eqnarray}}
\newcommand{\nn}{\nonumber}
\newcommand{\hc}{\mathrm{h.c.}}
\newcommand{\eps}{\epsilon}
\newcommand{\bwt}{\begin{widetext}}
\newcommand{\ewt}{\end{widetext}}
\newcommand{\draftnote}[1]{{\bf\color{blue} #1}}
\newcommand{\cO}{{\cal O}}
\newcommand{\cL}{{\cal L}}
\newcommand{\cM}{{\cal M}}
\newcommand{\fref}[1]{Fig.~\ref{fig:#1}} 
\newcommand{\eref}[1]{Eq.~\eqref{eq:#1}} 
\newcommand{\aref}[1]{Appendix~\ref{app:#1}}
\newcommand{\sref}[1]{Section~\ref{sec:#1}}
\newcommand{\tref}[1]{Table~\ref{tab:#1}}
\title{\LARGE{{\bf{Gravitational Wave Signals from Multiple Hidden Sectors} \\
}}}
\author{{\bf {Paul Archer-Smith, Dylan Linthorne, and Daniel Stolarski}}}
\affiliation{
Ottawa-Carleton  Institute  for  Physics,  Carleton  University,\\
1125  Colonel  By  Drive,  Ottawa,  Ontario  K1S  5B6,  Canada
}
\email{
Paul.Smith3@carleton.ca \\
Dylan.linthorne@carleton.ca \\
stolar@physics.carleton.ca
}
\begin{abstract}
We explore the possibility of detecting gravitational waves generated by first-order phase transitions in multiple dark sectors. $N$naturalness is taken as a sample model that features multiple additional sectors, many of which undergo phase transitions that produce gravitational waves. We examine the cosmological history of this framework and determine the gravitational wave profiles generated. These profiles are checked against projections of next-generation gravitational wave experiments, demonstrating that multiple hidden sectors can indeed produce unique gravitational wave signatures that will be probed by these future experiments. 
\end{abstract}
\maketitle

\section{Introduction} 
\label{sec:intro} 
The recent experimental detection of gravitational waves~\cite{Abbott:2016blz} gives humanity a new way to observe the universe. Future experiments~\cite{Danzmann:1994mma,Seto:2001qf,Crowder:2005nr,PhysRevD.72.083005,Harry:2006fi,Janssen:2014dka,Audley:2017drz,2017arXiv170200786A,Sato_2017,10.1093/ptep/pty078} will greatly expand the frequency range observable. Thus far, experiments have only observed recent events such as black hole mergers, but phase transitions in the early universe can leave an imprint as a \textit{stochastic} gravitational wave background~\cite{Witten:1984rs,Hogan:1984hx,Hogan:1986qda,PhysRevLett.65.3080,Caprini:2015zlo,Mazumdar:2018dfl}. Thus, searches for this background of gravitational waves can give direct information of the history of the universe before big bang nucleosynthesis. 
Because gravity is universal, gravitational waves can allow us to probe hidden sectors that couple very weakly, or not at all, to the Standard Model as long they are reheated after inflation. This was first explored in~\cite{Schwaller:2015tja}, and there has been significant work on this idea since~\cite{Jaeckel:2016jlh,Addazi:2016fbj,Hardy:2016mns,Dienes:2016vei,Tsumura:2017knk,Acharya:2017szw,Bernal:2017kxu,Aoki:2017aws,Heikinheimo:2018esa,Geller:2018mwu,Croon:2018erz,Baldes:2018emh,Bai:2018dxf,Breitbach:2018ddu,Fairbairn:2019xog,Helmboldt:2019pan,Caputo:2019wsd,Bertone:2019irm}. 

In this work, we explore the possibility of having multiple decoupled hidden sectors. Large numbers of hidden sectors can solve the hierarchy problem as in the Dvali Redi model~\cite{Dvali:2009ne}, in the more recently explored $N$naturalness~\cite{Arkani-Hamed:2016rle} framework, or in orbifold Higgs models~\cite{Craig:2014aea,Craig:2014roa}. They can also be motivated by dark matter considerations~\cite{Chialva_2013,Dienes:2011ja,Dienes:2011sa,Dienes:2016vei}. 
Motivated by solutions to the hierarchy problem, we consider hidden sectors with the same particle content as the Standard Model that have all dimensionless couplings (defined at some high scale) equal to those of the Standard Model. The only parameter that varies across sectors is the dimension-two Higgs mass squared parameter, $m_H^2$. This simple ansatz can lead to very rich phenomenology and interesting gravitational wave spectra, but we stress that it is only a starting point for exploring the space of theories with multiple hidden sectors.

In this setup, there are two qualitatively different kinds of sectors:
\begin{itemize}
\item \textbf{Standard Sectors}: Those with $m_H^2 < 0$ where electroweak symmetry is broken by the vacuum expectation value (vev) of a fundamental scalar. As in~\cite{Arkani-Hamed:2016rle}, we assume that the standard sector with the smallest absolute value of $m_H^2$ is the Standard Model.
\item \textbf{Exotic Sectors}: Those with  $m_H^2 > 0$. In this case, electroweak symmetry is preserved below the mass of the Higgs, and broken by the confinement of QCD~\cite{Susskind:1978ms}.
\end{itemize}
Cosmological observations, particularly limits on extra relativistic degrees of freedom at the time of Big Bang Nucleosynthesis and the time of the formation of the cosmic microwave background (CMB)~\cite{Aghanim:2018eyx}, require that most of the energy in the universe is in the Standard Model sector as we will quantify. Therefore, the hidden sectors cannot be in thermal equilibrium at any time, and the physics of reheating must dump energy preferentially in the Standard Model sector. This can be accomplished with primordial axionlike particle (ALP) models~\cite{MARSH20161,PhysRevLett.121.201303} and with the reheaton method~\cite{Arkani-Hamed:2016rle}. We will also explore alternative parameterizations of reheating that satisfy this condition.

In all the above models, there is some energy in the hidden sectors, and these sectors undergo thermal evolution independent of the SM sector. If their initial reheating temperature is above their weak scale, the standard sectors will undergo phase transitions associated with the breaking of electroweak symmetry and with confinement of QCD. The exotic sectors will also undergo a phase transition when QCD confines and electroweak symmetry is broken simultaneously. The condition for these transitions to leave imprints on the stochastic gravitational wave spectrum is that they strongly first-order phase transitions (SFOPT)~\cite{Witten:1984rs,Hogan:1984hx,Hogan:1986qda,PhysRevLett.65.3080}. This does not occur at either the electroweak or QCD phase transition in the SM, but as we will show, it does happen for the QCD phase transition in some standard sectors and in \textit{all} exotic sectors that reheat above the QCD phase transition. 

This work is organized as follows: section \ref{sec:nn} introduces the particle content of the model, section \ref{sec:dQCD} discusses the phase transition behaviour of both the standard and exotic sectors present, section \ref{sec:reheat} lays out hidden sector reheating, section \ref{sec:constraints} applies constraints from cosmological observables allowing for the calculation of gravitational wave signatures in section \ref{sec:gw}, and, finally, section \ref{sec:conclusion} ties everything up.

\section{Particle Setup}
\label{sec:nn}
We consider the following Lagrangian as in~\cite{Arkani-Hamed:2016rle}:
\begin{equation}
{\cal L} = \sum_{i=-N/2}^{N/2} {\cal L}_i,
\end{equation}
with ${\cal L}_0 = {\cal L}_{\rm SM}$ being the Standard Model Lagrangian, and ${\cal L}_i$ being a copy of the SM Lagrangian with different fields, but with all dimensionless parameters the same. Each of the Lagrangians does contain a dimensionful operator:
\begin{equation}
{\cal L}_i \subset  - \left(m_H^2\right)_i H_i^\dagger H_i  
\end{equation}
where $H_i$ is a Higgs field in each sector, and the mass term is parametrically given by
\begin{equation}\label{eqn:massParam}
\left(m_H^2\right)_i \sim -\frac{\Lambda_H^2}{N}(2i+r),
\end{equation}
where $\Lambda$ is some high-scale cutoff, $N$ is the number of sectors, and $r$ is the mass parameter in the SM in units of $\Lambda_H^2/N$. We view the parameterization of Eq.~(\ref{eqn:massParam}) as a random distribution in theory space up to the cutoff $\Lambda$: therefore, this setup solves the hierarchy problem if $r \sim \mathcal{O}(1)$~\cite{Arkani-Hamed:2016rle}\footnote{Constraints require $r$ to be somewhat smaller than 1.} and our sector is the one that that has the smallest absolute value of the Higgs mass parameter. We have taken for simplicity that there are equal numbers of sectors with positive and negative $m_H^2$, but this assumption does not affect our analysis. This $N$naturalness framework can be generalized: the various sectors can possess a wide range of particle content that can be freely selected by the model builder. The one exception to this is that ``our" sector must consist of the Standard Model.

From the above Lagrangians, the Higgs in sectors with $i \geq 0$ will get a VEV given by
\begin{equation}\label{eqn:vevs}
v^i = \sqrt{-(m_H^2)_i/\lambda_i} \sim \Lambda_H\sqrt{\frac{2i + r}{\lambda N}},
\end{equation}
$\lambda_i$ is the quartic coefficient of the scalar potential and is the same across all sectors, $\lambda_i=\lambda$. This is another way to see how this framework can solve the hierarchy problem: the Higgs VEV is parametrically smaller than the cutoff for $N \gg 1$. 
The ``standard sectors'' with $i>0$ feature electroweak symmetry breaking just like in the SM; however, the VEVs scale with the changing mass parameter: $v_i \sim v_{\rm SM}\sqrt{i}$. This means that the masses of the fermions and the $W$ and $Z$ will also increase proportional to $\sqrt{i}$. The consequences of this scaling on the confinement scale of QCD in the $i \geq 1$ sectors is further discussed in Sec.~\ref{sec:dQCD}. 

The ``exotic sectors'' with $i<0$ provide a radical departure from our own. $m_H^2 > 0$ leads to no VEV for the Higgs, and electroweak symmetry is only broken at very low scales due to the phase transition from free quarks to confinement at the QCD scale $\Lambda_{QCD}$~\cite{Susskind:1978ms}, and the masses of the $W$ and $Z$ are comparable to those of QCD resonances. The masses of fundamental fermions are produced via four-fermion interactions generated after integrating out the SU$(2)$ Higgs multiplet. This leads to very light fermions: 
\begin{equation}\label{eqn:exoticFM}
m_f \sim y_f y_t \Lambda_{QCD}^3/(m^2_H)_i \leq 100 \,\ev,
\end{equation}
with $y_f$ representing the Yukawa coupling to fermion $f$. As we will see, the extremely light quarks that appear in these sectors dramatically change the nature of the QCD phase transition --- unlike the SM, the transition is strongly first order. Again, this is further developed in Sec.~\ref{sec:dQCD}. Crucially, this results in the production of gravitational waves. This is the physical signature we explore in this paper; the calculation and results are presented in Sec.~\ref{sec:gw}.

\section{QCD Phase Transition}
\label{sec:dQCD}
%
We now study the nature of the QCD phase transition across the different sectors. Due to the confining nature of QCD, the exact nature of the phase transition is often difficult to ascertain analytically and requires the study of lattice simulations. In the SM, it is known that the phase transition is a crossover and does not lead to gravitational wave signals~\cite{Aoki:2006we,Bhattacharya:2014ara}. In the general case with three or more colours, the phase transition can be strongly first order in two regimes~\cite{SVETITSKY1982423,Pisarski:1983ms,Panero:2009tv}:
\begin{itemize}
\item three or more light flavours and
\item no light flavours. 
\end{itemize}

Light indicates a mass small compared to the confinement scale $\Lambda_{QCD}$, but what that means quantitatively is not precisely determined. In the SM, the up and down quarks are light, but the strange is not sufficiently light for an SFOPT.
%
For the standard sectors in our setup, the quark masses increase with increasing VEV, so for sufficiently large $i$, all the quarks will be heavier than $\Lambda_{QCD}$,\footnote{$\Lambda_{QCD}$ does vary with $i$, but the sensitivity is very weak as we will see below.} and those large $i$ sectors will undergo an SFOPT if they are reheated above the the confinement scale. 
Conversely, exotic sectors with zero VEV feature six very light quarks, so \textit{all} the exotic sectors undergo SFOPT at the temperature of QCD confinement. 

We now calculate the QCD confinement scale for each sector following the same procedure as \cite{Cui:2011wk}. First, due to the parameters of each sector being taken to be identical save for the Higgs mass squared (thus $v \neq v_i$, where $v$ is the SM VEV), we assume that the strong coupling of every sector is identical at some high scale. Using the one-loop running, the $\beta$ function can be solved:
\begin{equation}\label{eqn:QCDrunningi}
\alpha_{s}^i (\mu) = \frac{2\pi}{11-\frac{2n^i_f}{3}}\frac{1}{\ln{\mu/\Lambda^i}},
\end{equation}
where $n_f^i$ is the number of quark flavours with mass less than $\mu/2$ and $\Lambda^i$ is the scale where it would confine if all quarks remain massless. In the SM defined at scales well above all the quark masses, we have $\Lambda_{QCD} = 89 \pm 5$ MeV in $\overline{MS}$~\cite{PhysRevD.98.030001}. 
Because we have set the strong couplings equal at high scales, $\Lambda = \Lambda^i$ for all $i$ at high scales for all sectors. However, since the masses of the quarks in each sector are different, we end up with a unique running of the coupling for each sector. At every quark mass threshold for a given sector, we match the coupling strengths above and below the threshold and determine the new $\Lambda^i$ for the lower scale. For example, at the mass of the top quark, we match a five-flavour coupling with the six-flavour one:
\begin{equation}
\alpha_s^{i(5)}(2 m^i_t) = \alpha_s^{i(6)}(2 m^i_t) 
\end{equation} 
and thus
\begin{equation}
\Lambda_{(5)}^i = (m_t^{i})^{2/23}(\Lambda_{(6)}^i)^{21/23}.
\end{equation}
Suppressing the $i$'s for notational cleanliness, we can arrive at similar relations at the bottom and charm thresholds
\begin{equation}
\begin{split}
\Lambda_{(4)} = (m_b)^{2/25}(\Lambda_{(5)})^{23/25},
\\
\Lambda_{(3)} = (m_c)^{2/23}(\Lambda_{(4)})^{25/27}.
\end{split}
\end{equation}
These can be combined to show that
\begin{equation}
\Lambda_{(3)} = (m_t m_b m_c)^{2/27}(\Lambda_{(6)})^{21/27}.
\end{equation}
This type of matching procedure can be done as many times as necessary for a given sector. The process terminates when $\Lambda_i$ for a given scale is larger than the next quark mass threshold (i.e running the scale down arrives at the $\Lambda_{QCD}$ phase transition before reaching the next quark mass scale).
In cosmological terms, we can envision a sector's thermal history unfolding, whereas the plasma cools below each quark mass threshold and said quarks are frozen out. At a certain point, the sector arrives at the QCD phase transition and confinement occurs --- if this occurs when $\geq 3$ quarks are at a much lower scale or all quarks have already frozen out, we get the desired phase transition.   

\subsection{Standard Sectors}
As shown in Eq.~(\ref{eqn:vevs}), for standard sectors with increasing index $i$, the VEVs of said sectors increase $v_i\propto \sqrt{i}$. This leads to increasingly heavy particle spectra for higher sectors --- eventually leading to sectors that are essentially pure Yang-Mills that feature strong first-order phase transitions. This, of course, prompts the question: at what index $i$ do said phase transitions begin?
Using the methods outlined in the prior section we determine $\Lambda_{QCD}$ to have a relevant value of 
\begin{equation}\label{eqn:lambda2}
\Lambda^i_{(2)} = (m_s^i m_c^i m_b^i m_t^i)^{2/29}(\Lambda^i_{(6)})^{21/29}
\end{equation} 
at the energy scale we're interested in. $\Lambda^i_{(6)}$ is identical for all sectors and is taken to have a Standard Model value of $\Lambda^{(6)}_{MS} =(89 \pm 6)\, \mev$~\cite{PhysRevD.98.030001}. Rewriting Eq.~(\ref{eqn:lambda2}) in terms of Standard Model variables, 
\begin{equation}\label{eqn:lambda2adj}
\Lambda^i_{(2)} = (m_s m_c m_b m_t i^2)^{2/29}(\Lambda_{(6)})^{21/29}.
\end{equation}
where $m_q$ without a superscript is the mass of $q$ in the SM. We take the sector with SFOPT to be the ones when the mass of the up quark, down quark, and QCD phase transition scale are all comparable:
\begin{equation}
m^i_u \sim m_u \sqrt{i} \sim (m_s m_c m_b m_t i^2)^{2/29}(\Lambda_{(6)})^{21/29}.
\end{equation}
This can be solved for $i$:
\begin{equation}\label{eqn:critIndex}
i^c \sim \frac{(m_s m_c m_b m_t)^{4/21}(\Lambda_{(6)})^{2}}{( m_u)^{58/21}} \sim 10^6.
\end{equation}
As we will see in Sec.~\ref{sec:reheat}, in the original $N$naturalness setup~\cite{Arkani-Hamed:2016rle}, the energy dumped into the $i$th sector scales as $i^{-1}$, so there will not be enough energy in the sectors with $i>i^c$ to see a signature of these phase transitions. 
However, if we move away from the original $N$naturalness reheating mechanism and begin exploring mirror sectors with large VEVs and with relative energy densities $\rho_i / \rho_{SM} \sim 10\%$, a possibility allowed by current constraints, we can have sectors with relatively high dark QCD scales that produce detectable gravitational waves. From Eq.~(\ref{eqn:lambda2}) we can determine the confinement scale of an arbitrary mirror sector. If we take Higgs VEVs as high as the GUT scale $\sim 10^{16}$ GeV, then we can use Eq.~(\ref{eqn:lambda2adj}) to get confinement scales as high as $\sim 38 \,\gev$.  The signals of this sector and other test cases like it are explored in Sec.~\ref{sec:gw}.

\subsection{Exotic Sectors}
In every exotic sector the fermion masses are exceptionally light: their masses are generated by dimension six operators with the Higgs integrated out as shown in Eq.~(\ref{eqn:exoticFM}), and are therefore all below the confinement scale. The exotic sectors all have identical one-loop running of the QCD gauge coupling, and thus all have approximately the same confinement scale given by $\Lambda_{\rm ex} \sim 90 \,\mathrm{MeV}$. These sectors all have six light fermions, so a strong first order phase transition occurs for all exotic sectors at this temperature. 
The confinement of these sectors directly leads to the production of both baryons and mesons as we have the spontaneous breaking of SU$(6) \,\times$ SU$(6) \rightarrow\,$ SU$(6)$ and thus 35 pseudo-Goldstone bosons (pions). The masses obtained through the phase transition can be approximated through the use of a generalization of the Gell-Mann--Oakes--Renner relation~\cite{Gell-Mann,Schwartz:2013pla},
\begin{equation}\label{eqn:gmor}
m^2_{\pi} = \frac{V^3}{F^2_{\pi}}(m_u + m_d),
\end{equation}
where $V \sim \Lambda_{QCD}$, $F_\pi$ is the pion decay constant. One expects that within a given sector $F_\pi \sim V \sim \Lambda_{QCD}$ \cite{Schwartz:2013pla} and as exotic sectors have $\Lambda_{ex} \sim 90 \,\mev$ while the SM features $\Lambda_{QCD} = (332\pm17)\,\mev$ \cite{PhysRevD.98.030001} we expect at most $\mathcal{O}(1)$ difference in the $\sqrt{\frac{V^3}{F_\pi^2}}$ coefficient relative to the SM value. So, for pions in exotic sector $i$:
\begin{equation}
m_{\pi}^i \sim \sqrt{\frac{m_a^i+m_b^i}{m_u + m_d}} m_{\pi}.
\label{eq:ith_pion_mass}
\end{equation}
Here, $a$ and $b$ denote the component quark flavours.

\section{Reheating $N$ Sectors}
\label{sec:reheat}
A key issue within $N$naturalness is how to predominantly gift energy density to our own sector so as to not be immediately excluded by cosmological constraints, particularly those from number of effective neutrinos ($N_{eff}$). Here we review the results of~\cite{Arkani-Hamed:2016rle}. Reheating occurs through the introduction a ``reheaton'' field. After inflation, the reheaton field possesses the majority of the energy density of the Universe. Although this field can generically be either bosonic or fermionic, we reduce our scope to a scalar reheaton $\phi$. Our focus is primarily the production of gravitational waves from multiple sectors and a fermion reheaton does not change the scaling of the energy density of the exotic sectors and thus does not affect expected gravitational wave profiles.

In order to maintain the naturalness of our SM sector, the reheaton coupling is taken to be universal to every sector's Higgs. However, a large amount of the Universe's energy density must ultimately be deposited in our own sector for $N$naturalness to avoid instant exclusion. In order to accomplish this, the decay width of the reheaton into each sector must drop as $\vert m_H\vert$ grows. If we insist that the reheaton is a gauge singlet that is both the dominant coupling to every sector's Higgs and lighter than the naturalness cutoff $\Lambda_H/\sqrt{N}$, then we construct a model that behaves as desired. 
The appropriate Lagrangian for a scalar reheaton $\phi$ is: 
\begin{equation}\label{eqn:nnLagrangian}
\cL_\phi \supset -a\phi\sum_i\vert H_i\vert^2 - \frac{1}{2}m^2_\phi \phi^2.
\end{equation}
Note that cross-quartic couplings of the form $\kappa\vert H_i\vert^2\vert H_j\vert^2$ that could potentially ruin the spectrum of $N$naturalness are absent, taken to be suppressed by a very small coupling. Effective Lagrangians for the two different types of sectors present in this theory can be obtained by integrating out of the Higgs bosons in every sector:
\begin{equation}\label{eqn:nnEffLag}
\begin{split}
\cL_\phi^{v \neq 0} &\supset C_1 a y_q\frac{v}{m_h^2}\phi q q^c,
\\
\cL_\phi^{v = 0} &\supset C_2 a \frac{g^2}{16\pi^2 m_H^2}\phi W_{\mu\nu}W^{\mu\nu},
\end{split}
\end{equation}
with $C_i$ representing numerical coefficients, $g$ the weak coupling constant, and $W^{\mu\nu}$ the SU$(2)$ field strength tensor.
Immediately from Eq. (\ref{eqn:nnEffLag}), we can see that the matrix element for decays into standard sectors is inversely proportional to that sectors Higgs mass, $\cM_{m_H^2 < 0} \sim 1/m_{h_i}$ (since $v\sim m_H$). 
The loop decay of $\phi \rightarrow \gamma\gamma$ is always subleading and can be neglected. It should be noted that as one goes to sectors with larger and larger VEVs, the increasing mass of the fermions ($m_f \sim v_i \sim v_{SM}\sqrt{i}$) eventually leads to situations where the decay to two on-shell bottom or charm quarks is kinematically forbidden, $m_\phi < 2 m_q$. For sectors where this kinematic threshold is passed for charm quarks, the amount of energy in these sectors becomes so small that contributions to cosmological observables can be safely ignored. All in all, we end up with a decay width that scales as $\Gamma_{m_H^2<0} \sim 1/m_h^2$. Since we can expect energy density to be proportional to the decay width, $\frac{\rho_i}{\rho_{SM}} \approx \frac{\Gamma_i}{\Gamma_{SM}}$, this indicates that energy density of standard sectors falls:
\begin{equation}\label{eqn:edSS}
\rho_i \sim r_{s}\frac{\rho_{SM}}{i}
\end{equation} 
with $r_s$ being the ratio of the energy density of the first additional standard sector over the energy density of our sector.
For the exotic sectors, Eq.~(\ref{eqn:nnEffLag}) indicates a matrix element scaling $\cM_{m_H^2>0} \sim 1/m_{H_i}^2$ and is also loop suppressed. 
This leads to a significantly lower energy density than the standard sectors. Both the decay width and energy density for these sectors scale as
\begin{equation}
\Gamma_{m_H^2>0} \sim \rho_i \sim 1/m_H^4 \sim 1/i^2.
\label{eq:reheat_exotic}
\end{equation} 

As a final note, in this setup the reheating temperature of the SM, $T_{RH}$, has an upper bound on the order of the weak scale. If this bound is not observed, the SM Higgs mass would have large thermal corrections --- leading to the branching ratios into other sectors being problematically large~\citep{Arkani-Hamed:2016rle}. Thus we only consider relatively low reheating temperatures $\lesssim 100$ GeV.

Ultimately, after examining the gravitational wave case produced by standard $N$naturalness, we also consider a more generic parameterization where the reheating temperature of each sector is a free parameter and is in general uncorrelated with the Higgs mass parameter. This allows us to explore a broader model space with multiple dark sectors at a huge range of scales. For these models, the reheating mechanism remains unspecified.

\section{Constraints}
\label{sec:constraints}
In general, the multi-hidden-sector models explored feature a huge number of (nearly) massless degrees of freedom. Dark photons and dark neutrinos abound in these sectors and, assuming a relatively high reheat temperature, the leptons, quarks, and heavy bosons of these sectors can also be relativistic. In $N$naturalness this feature is realized quite dramatically: each of the $N$ sectors possess relativistic degrees of freedom. The presence of these particles can have two main effects: extra relativistic particles can alter the expansion history of the universe through changes to the energy density or hidden sectors can feature annihilations that reheat the photons or neutrinos of our sector near Big Bang Nucleosynthesis (BBN) and affect the light element abundances. The effective number of neutrino species, $N_{eff}$, is impacted by these contributions and, as such, is the strictest constraint that must be dealt with when studying these type of multi-phase-transition models. 
The SM predicts that $N^{SM}_{eff} = 3.046$ \cite{Mangano:2005cc}. This is in good agreement with the $2\sigma$ bounds from studies of the Cosmic Microwave Background (CMB) by Planck combined with baryon acoustic oscillations (BAO) measurements \cite{Aghanim:2018eyx}:
\begin{equation}\label{eqn:neffBounds}
N_{eff} = 2.99^{+0.34}_{-0.33}.
\end{equation}
Various different assumptions about the history of the universe can be made and different data sets can be chosen to obtain slightly different results \cite{Breitbach:2018ddu} --- for the purposes of this exploratory work, wading through this landscape is unnecessary. Additionally, 
\begin{equation}
\frac{(\Delta N^i_{eff})_{CMB}}{(\Delta N^i_{eff})_{BBN}} \geq 1
\end{equation}
for any decoupled hidden sector~\cite{Arkani-Hamed:2016rle}. Because the constraints on $N_{eff}$ are stronger at photon decoupling than at BBN, we can focus purely on the constraints provided by the former.
Future CMB experiments~\cite{Abazajian:2016yjj} will improve the bound from Eq.~(\ref{eqn:neffBounds}) by about an order of magnitude. This could significantly reduce the allowed temperature ratio of any hidden sector or, alternatively, could provide evidence for such sectors in a way that is complementary to the gravitational wave signatures described below. 
For fully decoupled sectors that never enter (or reenter) thermal equilibrium with our sector, we obtain additional contributions to $N^{SM}_{eff}$ \cite{Breitbach:2018ddu}
\begin{equation}\label{eqn:DeltaNeff1hs}
\Delta N_{eff} = \frac{4}{7}\left(\frac{11}{4}\right)^{4/3}g_h \xi_h^4.
\end{equation} 
Here, $g_h$ represents the effective number of relativistic degrees of freedom for the hidden sector\footnote{$g_h = N_{\rm boson} + 7 N_{\rm fermion}/8$.}, and we parameterize the hidden sector temperature by~\cite{Breitbach:2018ddu}
\begin{equation}
\xi_h \equiv  \frac{T_{h}}{T_{\gamma}},
\label{eq:xi}
\end{equation}
and these should be evaluated at the time of photon decoupling. 
%
%
We take this approach and generalize it to include many additional sectors:
\begin{equation}\label{eqn:DeltaNeff}
\Delta N_{eff} = \sum_i \frac{4}{7}\left(\frac{11}{4}\right)^{4/3}g_{i} \xi^4_{i}.
\end{equation} 
For a dark sector with one relativistic degree of freedom, its temperature must be $T_{DS}\sim 0.6\, T_{\rm SM}$ to not be excluded. Applying the energy density formula \cite{Trodden:2004st},
\begin{equation}
\rho_i = \frac{\pi^2}{30}g_i T_i^4,
\end{equation}
to both said dark sector and the SM and then taking the ratio indicates that the dark sector would have an energy density $\rho \sim 0.038\, \rho_{\rm SM}$. 

\subsection{Exotic Sector Contributions}
We begin by computing the constraints on exotic sectors; these are significantly weaker than those for standard sectors~\cite{Arkani-Hamed:2016rle}. 
At the time of photon decoupling, $T_\gamma \sim 0.39 \;\ev$ while the temperature of the exotic sectors is lower. This means that for sectors with small and moderate $i$, we can use Eqs.~(\ref{eqn:exoticFM}) and~(\ref{eq:ith_pion_mass}) to see that the pions will be nonrelativistic  leaving at most $7.25$ effective degrees of freedom per sector from photons and neutrinos.  For very large $i$, the pions can be much lighter, but those sectors also have very little energy in them in the standard reheating scenario. 
Coupling the number of effective degrees of freedom per sector with the energy density scaling of $\sim 1/m_H^4$ as in Eq.~(\ref{eq:reheat_exotic}) means that the zero VEV sectors have small temperature ratios. Assuming a reheating temperature of $100$ GeV and a completely uniform distribution of sectors, the temperature of the first exotic sector is slightly more than $6\%$ of our sector at reheating. Applying Eq.~(\ref{eqn:DeltaNeff}) to this particular situation gives us:
\begin{equation}
\Delta N_{eff} = \sum_i \frac{4}{7}\left(\frac{11}{4}\right)^{4/3}g_{i} \left(\frac{(T_{RH_{E1}}/T_{RH})}{i^{1/2}}\right)^4 \sim 10^{-4},
\end{equation}     
with $T_{RH_{E1}}/T_{RH}$ being the ratio of the reheat temperatures of the first exotic sector and our own sector ($0.06$ in standard $N$naturalness with $r = 1$).  This sum is dominated by $i=1$; the sector with the lowest Higgs mass (and thus the most energy density) gives us a contribution of $\mathcal{O} (10^{-4})$ to $\Delta N_{eff}$. Evolving the sector thermal histories forward in time to the recombination era gives us a slightly larger value, but still of order $\mathcal{O}(10^{-4})$, well below current CMB bounds. 
It should be noted that modifying the exotic sectors' structure (e.g. adjusting the exotic sectors to have a lower Higgs mass squared or clustering multiple hidden sectors close to the first exotic one) leads to a $\Delta N_{eff}$ contribution that is larger than the base $N$naturalness case. This increase is typically not excluded by current bounds, indicating a large degree of liberty in the structure and number of exotic hidden sectors.

\subsection{Standard Sector Contributions}

Within the context of vanilla $N$naturalness, the majority of contributions arise from standard sectors. This is explored in detail in \cite{Arkani-Hamed:2016rle}; here we briefly summarize these arguments. All additional standard sectors are very similar to our own: they have the same particle content and couplings and differ only by the Higgs mass. As our sector is taken to be the lightest so as to be preferentially reheated, every other standard sector features an earlier freeze-out of their respective particles. This ultimately leads to each sector having at most the same number of relativistic degrees of freedom as the SM.

In \cite{Arkani-Hamed:2016rle}, the standard sector contributions are expressed as:
\begin{equation}
\Delta N_{eff} = \frac{1}{\rho^{us}_\nu} \sum_{i\neq us} \rho_i.
\end{equation}
In the case that the reheaton is lighter than the lightest Higgs (ours), this can be expressed as
\begin{equation}
\begin{split}
\Delta N_{eff} &\sim \sum^{N_b}_{i=1}\frac{1}{2i+1}+\frac{y_c^2}{y_b^2}\sum^{N_c}_{i=N_b + 1}\frac{1}{2i+1}\\
& \simeq \frac{1}{2}\left(\log 2N_b + \frac{y_c^2}{y_b^2}\log \frac{N_c}{N_b}\right)
\end{split}
\end{equation}
with $y_{c,b}$ representing the charm and bottom Yukawa couplings, respectively, and 
\begin{equation}
N_{b,c} = \left(\frac{m_\phi^2}{8m^2_{b,c}}-\frac{1}{2}\right)
\end{equation}
with $m_\phi$ being the mass of the reheaton.

Application of these results indicates that for a majority of the parameter space, vanilla $N$naturalness requires mild fine-tuning ($r$ in Eq.~(\ref{eqn:vevs}) set to a value $\lesssim 1$). Numerical results for the fine-tuning required for various reheaton masses were presented in \cite{Arkani-Hamed:2016rle}.

\subsection{Generalized Reheating Scenarios}\label{sec:genReheat}

The generalization of possible reheating mechanisms mentioned in section~\ref{sec:reheat} --- where the reheating mechanism no longer depends on the Higgs' mass parameter of a given sector --- opens up a wide range of hidden sectors for study. Specifically, this allows mirror sectors with large Higgs VEVs to be reheated to significant energy densities and thus produce gravitational waves with enough power to be detected. Crucially, despite this analysis being limited to mirror sectors with large Higgs masses, this analysis pertains to any strong, confining phase transition at high scales.
Since $N_{eff}$ constraints remain our strongest cosmological bounds for massive standard sectors, our starting point for exploring the limits of high transition temperatures is Eq.~(\ref{eqn:DeltaNeff}). Assuming heavy, standard sectors (with the only relativistic particles being photons and neutrinos) we can saturate the bounds of Eq.~(\ref{eqn:neffBounds}) and solve for the maximum temperature allowed for any number of sectors:
\begin{equation}\label{eqn:energyDensityAllowed}
\begin{split}
T_i &\sim 0.38 \,T_{SM} \,\,\,\,\, \mathrm{1}\,\, \mathrm{hidden}\,\, \mathrm{sector},
\\
T_i &\sim 0.25 \,T_{SM} \,\,\,\,\, \mathrm{5} \,\,\mathrm{hidden}\,\, \mathrm{sectors},
\\
T_i &\sim 0.21 \,T_{SM} \,\,\,\,\, \mathrm{10} \,\,\mathrm{hidden}\,\, \mathrm{sectors},
\\
T_i &\sim 0.12 \,T_{SM} \,\,\,\,\, \mathrm{100} \,\,\mathrm{hidden}\,\, \mathrm{sectors},
\end{split}
\end{equation}
where all the hidden sectors have the same temperature as one another.

Using these restrictions, we can examine the behaviour of standard sectors with a much larger VEV than our own. In terms of the $N$naturalness framework, this means we can get an SFOPT for QCD if we look at sectors with $i$ greater than the critical index of Eq.~(\ref{eqn:critIndex}) where all the quark masses are above the QCD confinement scale, as long as their temperatures are below the bounds presented here. 

\section{Gravitational Wave Signals}
\label{sec:gw}
We now turn to the gravitational wave signatures of our setup. At high temperatures, each of the hidden sectors has QCD in the quark/gluon phase, but at temperatures around $\Lambda_{\rm QCD,i}$, the $i^{\mathrm{th}}$ sector undergoes a phase transition into the hadronic phase that we computed for the different sectors in Sec.~\ref{sec:dQCD}. As discussed in that section, this phase transition will be strongly first order (SFOPT) for certain numbers of light quarks, which will generate gravitational waves. This differs from QCD in the SM sector, as the PT is a crossover and not first order~\cite{Fodor:2001pe}. 
A SFOPT proceeds through bubble nucleation, where bubbles of the hadronic phase form in the vacuum of the quark phase. These bubbles will expand, eventually colliding and merging until the entire sector is within the new phase.  These bubbles are described by the following Euclidean action~\cite{Linde:1981zj}:
\begin{equation}\label{eqn:EuclideanAction}
S_{E}(T) = \frac{1}{T}\int d^3x \bigg[\frac{1}{2}(\nabla\phi)^2 + V(\phi,T)  \bigg],
\end{equation}
where the time component has been integrated out due to nucleation occurring not in vacuum but in a finite temperature plasma. $\phi$ is the symmetry-breaking scalar field with a nonzero VEV. In the case of the chiral phase transition, the scalar field breaking the SU$(N_f)_{R}$ $ \times$ SU$(N_f)_{L}$ chiral symmetry is the effective quark condensate $\phi_{i} \sim \langle q\bar{q} \rangle_{i}$ of the respective sector. We leave the thermalized potential $V(\phi, T)$ general. As previously stated, an exact QCD potential at the time of the chiral phase transition is not well understood outside of lattice results. In~\cite{Bai:2018dxf} chiral effective Lagrangian was used to calculate a low-energy thermalized potential for confining SU$(N)$. 
The amount of energy density dumped into the individual sectors dictates the energy budget for the PT and hence for the gravitational waves. Assuming that the SM sector is radiation dominated, a quantity that characterizes the strength of the PT is the ratio of the latent heat of the phase transition, $\epsilon$, to the energy density of radiation, at the time of nucleation~\citep{Espinosa:2010hh},
\begin{equation}
\alpha     \equiv  \frac{\epsilon}{g_{*} \pi^2 (T^{nuc}_{\gamma})^4/30},
\label{eq:alpha}
\end{equation}
with $\epsilon$ being calculable from the scalar potential. Assuming that there is a negligible amount of energy being dumped back into the SM, which would cause significant reheating of $\rho_{\gamma}$,  the latent heat $\epsilon$ should correspond to the energy density of the hidden sector going through the PT. The parameter $g_{*}$ in the denominator of Eq.~(\ref{eq:alpha}) is the number of relativistic degrees of freedom at the time of the phase transition, with contributions from species in both the visible and dark sectors. It has weak temperature dependence in a single sector, but when dealing with multiple hidden sectors, $g_{*}$ gains contributions from all $N$ sector's relativistic degrees of freedom, weighted by their respective energy densities
\begin{equation}\label{eqn:RelaDOF}
g_{*} = g_{*,\gamma} + \sum_{i} g_{*,i} (\xi_{i})^4,
\end{equation}
with $\xi$ being the temperature ratio defined in Eq.~(\ref{eq:xi}). The bounds from effective number of neutrinos~\citep{Aghanim:2018eyx} mean that $\xi_i \lesssim 1$ for all $i$, so $g_{*} \approx g_{*,\gamma}$. In the case of dark QCD-like chiral  phase transitions, the temperature of the phase transition is on the order of the symmetry-breaking scale of the respective sector, $T_{h}^{i} \sim \mathcal{O}(\Lambda_{QCD,i})$. The work of~\citep{Bai:2018dxf} calculated $\alpha$ with an effective chiral Lagrangian has found various upper bounds. We take the optimistic scenario where the numerator is bounded above by the symmetry breaking scale
\begin{equation}\label{eqn:alpha2}
\alpha_i \approx \xi_i^{4} \approx \bigg( \frac{\Lambda_{QCD,i}}{T_{\gamma}^{nuc}}\bigg)^4,
\end{equation}
where $T_{\gamma}^{nuc}$ is the temperature of the SM photon bath at the time of the phase transition. Another important parameter to characterize the phase transition is its inverse timescale $\beta$~\citep{Caprini:2015zlo}. The inverse timescale can be calculated using the action in Eq.~(\ref{eqn:EuclideanAction}): 
\begin{equation}
\beta  \equiv \frac{dS_{E}(T)}{dt}\bigg|_{t = t_{nuc}}.
\end{equation}
The ratio of $\beta$ and the Hubble constant, at the time of nucleation, $H$ controls the strength of the gravitational wave (GW) signal,
\begin{equation}\label{eqn:BH}
\frac{\beta}{H} = T^{nuc}_h \frac{dS_{E}(T)}{dT}\bigg|_{T = T^{nuc}_h}.
\end{equation}
Due to the lack of a general analytic QCD potential, it is not possible to use Eq.~(\ref{eqn:BH}) to calculate $\beta/H$. There are dimensional arguments~\cite{Hogan:1984hx,Hogan:1986qda} that predict $\beta/H \sim 4 \textrm{Log}(M_{p}/ \Lambda_{QCD,i})$, although these arguments make specific assumptions about the potential. In more recent work, some authors~\citep{Helmboldt:2019pan,Bai:2018dxf} have attempted to estimate it using first-order chiral effective theories and the Polyakov-Nambu-Jona-Lasinio models which motivates a $\beta/H$ of $\mathcal{O}(10^4)$. These studies claim a large range of values with no consensus reached on the precise order of the scaled inverse timescale.  Under these circumstances, our signal projections will consider both extremes of the parameter space where, $\beta/H \sim 10 - 10^4$. A more realistic scenario may exist in between both cases.

\subsection{Production of Gravitational Waves}
\label{sec:signals}

Gravitational waves are produced with contributions from different components of the SFOPT's evolution.  It is commonplace to parameterize the spectral energy density in gravitational waves by~\citep{PhysRevD.75.043507} 
\begin{equation}\label{eqn:SpectralEng}
\Omega_{\textrm{GW}} (f) \equiv \frac{1}{ \rho_{c}} \frac{d \rho_{\textrm{GW}}(f) }{d \textrm{\,log}(f)} , 
\end{equation}
where $\rho_{c} = 3H^2/(8 \pi G)$ is the critical energy density. The total gravitational wave signal is a linear combination of three leading contributions:
\begin{equation}
h^2\Omega_{\textrm{GW}} \approx h^2\Omega_{\phi} + h^2\Omega_{v} + h^2\Omega_{turb} .
\end{equation}
Each components is scaled by its own unique efficiency factor, $\kappa$. The three leading-order contributions to the GW power spectrum are as follows:
\begin{itemize}
\item \textbf{Scalar field contributions $\Omega_{\phi}$}: Caused by collisions of the bubble walls, the solutions being completely dependent on the scalar field configuration, with efficiency factor $\kappa_{\phi} = 1 - \alpha_{\infty}/\alpha$~\citep{PhysRevD.45.4514, Huber_2008}. 
\item \textbf{Sound wave contributions $\Omega_{v}$}: Sound waves within the plasma after bubble collision will produce $\beta/H$ enhanced gravitational waves, with efficiency factor $\kappa_{v} \propto \alpha_{\infty}/\alpha$~\citep{PhysRevLett.112.041301}.
\item \textbf{Magnetohydrodynamical contributions $\Omega_{B}$}: Turbulence within the plasma, left over from the sound wave propagation, will produce gravitational waves with efficiency factor $\kappa_{
turb} \approx 0.1 \kappa_{v} $~\citep{PhysRevD.74.063521}.
\end{itemize}
The parameter $\alpha_{\infty}$ denotes the dividing line between the runaway regime  $(\alpha >\alpha_{\infty})$ and the nonrunaway regime $(\alpha <\alpha_{\infty})$. Explicitly \cite{Breitbach:2018ddu, Caprini:2015zlo, Espinosa:2010hh}, 
\begin{equation}\label{eqn:critPTstrength}
\alpha_{\infty} = \frac{(T^{nuc}_h)^2}{\rho_R}\left[\sum_{bosons} n_i\frac{\Delta m^2_i}{24} + \sum_{fermions} n_i\frac{\Delta m_i^2}{48}\right],
\end{equation} 
for particles with $n_i$ degrees of freedom that obtain mass through the phase transition. 

The exotic sectors have essentially massless degrees of freedom pre phase transition and pions with negligible masses post phase transition. Other composite particles, such as baryons, do gain a mass of the order of $\Lambda_{ex}$; this is, however, still much smaller than the order of $\rho_R$ leading to small $\alpha_{\infty}$ according to Eq.~(\ref{eqn:critPTstrength}).\footnote {It should also be noted that although free quarks cease to exist post phase transition in these exotic sectors, their masses are so light that they do not contribute relevant amounts to $\alpha_{\infty}$.} Heavy standard sectors that undergo SFOPT for QCD feature no baryons due to all quarks being above their respective QCD scales. They do, however, feature glueballs that obtain a mass of the order of the SFOPT and, just as in the case for exotic sectors above, feature small $\alpha_{\infty}$. 
 
Each component of the spectral energy density in Eq.~(\ref{eqn:SpectralEng}) is proportional to a power of their respective efficiency factors $\kappa$. The relative strength of the efficiency factors is dependent on the ratio $\frac{\alpha_{\infty}}{\alpha}$ --- since both $\alpha_{\infty}$ and $\alpha$ are parameterically small \cite{Bai:2018dxf,Helmboldt:2019pan}, a range of possible scenarios can occur. Here, we discuss the two ends of this spectrum: pure runaway walls and pure nonrunaway walls. The former scenario with runaway bubble walls leads to the efficiency factors for the sound wave and magnetohydrodynamics (MHD) contributions being small and ensures GWs are dominantly produced from bubble collisions, $h^2\Omega_{\textrm{GW}} \approx h^2\Omega_{\phi}$: this is what we assume for the remainder of this section. In the latter case, bubbles are nonrunaway (but $v_w \sim 1$ still \cite{Breitbach:2018ddu,Bodeker2017}) such that nonbubble collision contributions being important --- ultimately leading to significant changes to the GW profile. This case and the gravitational waves it produces are examined in the Appendix. Intermediate results are of course possible and would feature profiles somewhere in between the two extremes.

The form of the GW energy density at the time of nucleation is given by~\citep{Breitbach:2018ddu}
\begin{equation}\label{eqn::GWengdensity}
h^2\Omega_{\textrm{GW}}^* = 7.7\times 10^{-2} \bigg( \frac{\kappa_{\phi} \alpha}{1 + \alpha} \bigg)^2 \bigg( \frac{H}{\beta} \bigg)^2  S(f)
\end{equation}
where we use $v = 1$ for runaway bubbles. Quantities such as $\Omega_{\textrm{GW}}^*$ that are calculated at the time of nucleation are denoted with an asterisk, and they must then be evolved to relate to their values at the time of observation.  $S(f)$ is the spectral shape function for the signal and a parametric from has been found through numerical simulations~\cite{Huber_2008} of bubble wall collisions:
\begin{equation}\label{eqn::spectralshape}
S(f) = \frac{3.8 \,(f/f_p)^{2.8}}{1 + 2.8\,(f/f_p)^{3.8}}.
\end{equation}
The peak frequency $f_{p}$ is a function of the temperature of the SM at the time of nucleation.  The various hidden sectors can phase transition at different scales and therefore temperatures, causing a shift in the GW spectrum's peak frequency given by~\cite{Huber_2008}
\begin{equation}\label{eqn:peakFrequency}
f_{p} = 3.8 \times 10^{-8} \, \textrm{Hz}\, \bigg( \frac{\beta}{H}\bigg)\bigg(\frac{T_{\gamma}}{100 \, \textrm{GeV}}\bigg)\bigg(\frac{g_{*}}{100}\bigg)^{\frac{1}{6}},
\end{equation}
where $g_*$  is calculated using Eq.~(\ref{eqn:RelaDOF}), although, due to the lack of substantial reheating into the hidden sectors, the SM contribution is dominant. 

Now that the framework has been laid out for the creation of GW from a single SFOPT, we generalize to multiple sectors going under independent, coherent SFOPT. In the models presented in this paper, we consider a subset of $N$ hidden sectors that undergo a phase transition at a SM temperature of $T_{\gamma}^{i}$. As the GWs propagate in free space, the energy density and frequency spectrum, at the time of production $\Omega_{\textrm{GW}}^{*}(f)$, will redshift to today's value $\Omega_{\textrm{GW}}^{0}(f) = \mathcal{A}\, \Omega_{\textrm{GW}}^{*}((a_{0}/a) f) $. The redshifting factor $\mathcal{A}$ accounts for the redshifting of both $\rho_{\textrm{GW}} $ and $\rho_{c}$~\cite{Breitbach:2018ddu,PhysRevD.49.2837},
\begin{equation}\label{eqn::redshifting}
\mathcal{A} \equiv \bigg( \frac{a}{a_{0}} \bigg)^4 \bigg( \frac{H}{H_{0}}\bigg)^2
\end{equation}
where $a$ ($a_{0}$)  and $H$ ($H_{0}$) are the scale factor and Hubble constant at the time of nucleation (observation), respectively. Assuming that the sectors are completely decoupled before and after their respective SFOPT, the total GW signal that would be measured today is given by the coherent sum
\begin{equation}
\Omega_{\textrm{GW}} =  \sum_{i}^{N} \mathcal{A}^{i}\, \Omega_{\textrm{GW}}^{i,*}((a_{0}/a)_{i} f) .
\end{equation}
We assume that the parameters of the SFOPT do not differ between sectors: the relativistic degrees of freedom, phase transition rate, and the dark QCD scale, are all similar. This makes the redshifting factor $\mathcal{A}^i$ independent of sector number.
\begin{figure}[h!]
\centering
\includegraphics[scale=.3]{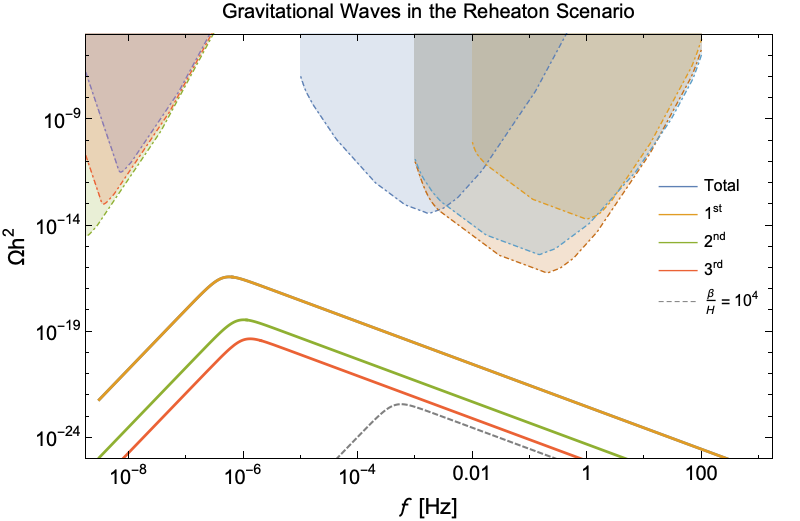} 
\caption{ Gravitational wave spectral energy density (solid curves) for standard $N$naturalness using the scalar reheaton model of section~\ref{sec:reheat}. The curve corresponding to the sum of the sectors is approximately equal to the $i=1$ curve. All contributions are assumed to be purely from bubble collisions $\Omega_{\phi}$. The colored solid lines use $\beta/H  = 10$ whereas the dashed gray line is the total contribution of all sectors for $\beta/H  = 10^4$ (the sum of all sectors is roughly equal to the $i = 1$ curve and sectors beyond the first are below the range of this plot).  The shaded dashed curves are the power law noise curves~\cite{PhysRevD.88.124032} calculated from expected sensitivity as described in Section~\ref{sec:detection}. The ones on the right are space-based interferometers: LISA~\citep{Audley:2017drz} (blue), DECIGO~\citep{10.1093/ptep/pty078} (light blue), BBO~\citep{PhysRevD.72.083005} (red). The ones on the left are for the pulsar timing array SKA~\cite{Janssen:2014dka} for exposure time of 5-years (purple), 10-years (orange), and 20-years (green). }
\label{fig::Nnatural}
\end{figure}
Applying this to the standard reheating scenario of $N$naturalness, introduced in~\ref{sec:reheat}, we get GW signals as seen in Fig.~\ref{fig::Nnatural}. Plotted are the individual contributions to the signal from each phase transitioned sector, as well as the coherent sum of all sectors. Future GW interferometers and pulsar timing array sensitivity curves are shown in comparison to the signal. The sensitivity curves are interpreted as the region of possible detection if intersected with the GW signal, and the construction of these curves is detailed in Section~\ref{sec:detection}. Notice that the total signal is dominated by the first sector's contribution. This is caused by the quartic temperature ratio suppression in Eq.~(\ref{eq:alpha}) and the large temperature gaps between adjacent sectors. Such a suppression leads to standard $N$naturalness evading future detector thresholds by a few orders of magnitude in units of energy density. 

This is not the case if we consider more generalized reheating scenarios. Once the restriction that sectors with small Higgs masses are preferentially reheated has been lifted, we can explore a much more vast landscape of hidden sectors than are allowed in the reheaton case. Here, we construct several different scenarios that are both detectable and demonstrate a variety of gravitational wave profiles. Specifically, we explore benchmarks that lead to a deviation in the peak behaviour of the total GW signal (the superposition of stochastic GW from individual SFOPT) from a standard power law signal.

It should be noted that the key phenomenological constraint on all of these models is $\Delta N_{eff}$, giving us a maximum allowed temperature ratio (when compared to the SM) for each reheated hidden sector: Eq.~(\ref{eqn:energyDensityAllowed}) shows the maximum temperature ratios for specific numbers of additional hidden sectors. Due to the rather harsh scaling of the GW strength $\alpha$ with the temperature ratio shown in Eq.~(\ref{eqn:alpha2}), we take the optimistic approach of keeping the temperature ratio as high as allowed by CMB data for all of the hidden sectors.

In the following, we focus on heavy standard sectors --- pure Yang-Mills sectors with much heavier particles (specifically quarks) and, as shown in Sec.~\ref{sec:dQCD}, the SFOPT these entail. The reason for this arises from Eq.~(\ref{eqn:peakFrequency}): every exotic sector features a phase transition that occurs at $\Lambda_{ex} \sim 90$ MeV. If we maximize the allowed temperature ratio, this gives us a (SM) photon temperature $T_{\gamma}$ that places our signal directly in the frequency void between the detection region of pulsar timing arrays and space-based interferometers (see Sec.~\ref{sec:detection}). The location of the peak can be changed by dropping the temperature ratio, but the adjustment required to end up with a signal with an appropriate peak frequency makes the overall signal too weak to detect. As shown in Sec.~\ref{sec:dQCD}, standard sectors can have much higher temperature phase transitions. As such, maintaining the maximum allowed temperature ratio between the hidden sector(s) and the SM gives a much larger photon temperature and a proportionally larger peak frequency; ultimately allowing for detection by space-based interferometers.

\begin{table*}[t]
\begin{center}
\begin{tabular}{ |p{2cm}||p{3cm}|p{3cm}|p{3cm}|p{2cm}|  }
 \hline
 \multicolumn{5}{|c|}{Parameters for multi-hidden sector benchmarks} \\
 \hline
\multicolumn{5}{|c|}{Maximized signal} \\
 \hline
 Sector & Higgs VEV (GeV) & $\Lambda^{AS}_{QCD}$ (GeV) & $T_\gamma$ (GeV) & Index\\
 \hline
 1 & $24.6 \times 10^{9}$ & 38.6 & 87.7 & $10^{16}$ \\
 \hline
\multicolumn{5}{|c|}{Large split} \\
 \hline
 Sector & Higgs VEV (GeV) & $\Lambda^{AS}_{QCD}$ (GeV) & $T_\gamma$ (GeV) & Index\\
 \hline
 1 & $246 \times 10^6$  & 10.8 & 30.3 & $10^{12}$ \\
 2 & $7.8 \times 10^9$  & 28.1 & 78.6 & $10^{15}$ \\
 \hline
 \multicolumn{5}{|c|}{Medium split} \\
 \hline
 Sector & Higgs VEV (GeV) & $\Lambda^{AS}_{QCD}$ (GeV) & $T_\gamma$ (GeV) & Index\\
 \hline
 1 & $246 \times 10^{6}$ & 10.8 & 30.3 & $10^{12}$ \\
 2 & $778 \times 10^6$  & 14.9 & 41.6 & $10^{13}$ \\
 \hline
 \multicolumn{5}{|c|}{Five sector} \\
 \hline
 Sector & Higgs VEV (GeV) & $\Lambda^{AS}_{QCD}$ (GeV) & $T_\gamma$ (GeV) & Index\\
 \hline
 1 & $246 \times 10^6$ & 10.8 & 38.8 & $10^{12}$ \\
 2 & $426 \times 10^6$ & 12.6 & 45.2 & $3\times10^{12}$ \\
 3 & $778 \times 10^6$ & 14.9 & 53.3 & $10^{13}$ \\
 4 & $1.3 \times 10^9$ & 17.3 & 62.1 & $3\times10^{13}$ \\
 5 & $2.5 \times 10^9$ & 20.5 & 73.3 & $10^{14}$ \\
 \hline
\end{tabular}
\caption{Outline of parameters used for the various multi-hidden-sector scenarios. The Higgs VEV is the VEV for the given additional sector, $\Lambda^{AS}_{QCD}$ is the QCD phase transition in the additional sector, and $T_\gamma$ is the temperature of the SM photon bath when the SFOPT occurs in the additional sector. The index indicates the equivalent sector from the $N$naturalness model (Eq.~(\ref{eqn:vevs})). It should be noted that although the various sectors undergo phase transitions at different temperatures, they are all assumed to be reheated to the same initial temperature.}\label{tab:cases}
\end{center}
\end{table*}

There are four scenarios that we examine, with key parameters presented in Table \ref{tab:cases}. 
\begin{itemize}
\item \textbf{Maximized signal:} A single additional heavy hidden sector reheated to a temperature that saturates current experimental bounds. The SM photon bath temperature at the time of the hidden sector PT is $87$ GeV. In the $N$naturalness framework this is equivalent to reheating a standard sector with $i\sim 10^{16}$ up to the maximum allowed temperature ratio.

\item \textbf{Large split scenario:} A scenario where two additional hidden sectors have been reheated --- these sectors have Higgs VEVs that are split by a factor of 
\begin{equation}
 \frac{v_{h1}}{v_{h2}} = \sqrt{10^3}.
\end{equation}
This results in a difference in the scale of the SFOPTs leading to the SM photon bath temperature changing a large amount during the time between the PTs. This, in turn, leads to a large separation in the peak frequency of their gravitational wave signals.  In the $N$naturalness framework this is equivalent to reheating two standard sectors, one with $i\sim 10^{12}$ and another with $i\sim 10^{15}$ up to the maximum allowed temperature ratio.

\item \textbf{Medium split scenario:} Similar to the previous case: these sectors have Higgs VEVs that are split by a factor of 
 \begin{equation}
 \frac{v_{h1}}{v_{h2}} = \sqrt{10},
\end{equation}
resulting in a much smaller difference in the peak frequency of their gravitational wave signals. In the $N$naturalness framework this is equivalent to reheating two standard sectors, one with $i\sim 10^{12}$ and another with $i\sim 10^{13}$ up to the maximum allowed temperature ratio.

\item \textbf{Five sector scenario:} Five sectors are reheated to the maximum allowed temperature ratio, each with VEVs that are 
\begin{equation}
 (v_{hi})/(v_{h(i+1)}) \sim \sqrt{3} 
\end{equation}
 larger than the previous sector.

\end{itemize}
In all cases where multiple sectors are reheated, we assume for simplicity that all the hidden sectors are reheated to the same temperature. 

\begin{figure*}[tb]
\centering
\begin{minipage}[c]{\textwidth}
\includegraphics[width=.45\textwidth ]{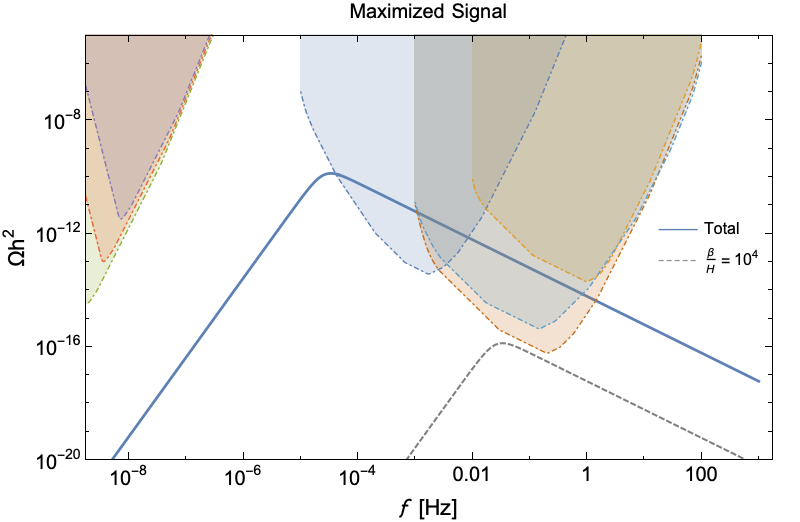}
\hfill
\includegraphics[width=.45\textwidth]{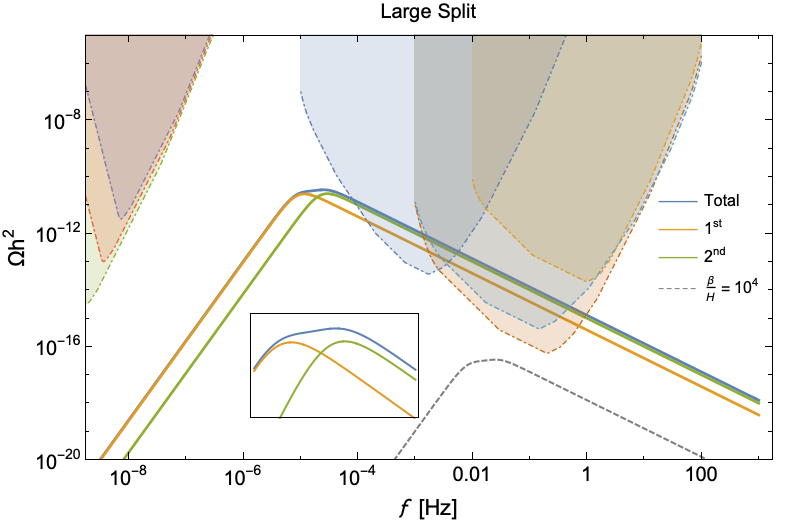} 
\hfill
\includegraphics[width=.45\textwidth]{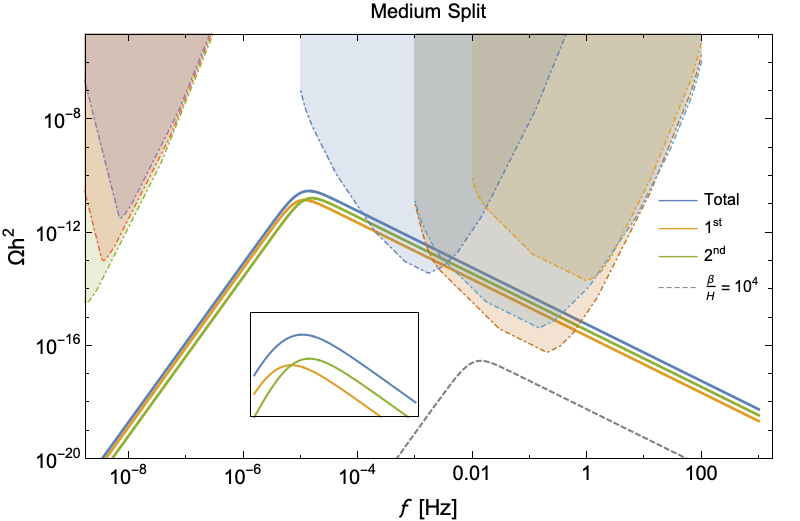} 
\hfill
\includegraphics[width=.45\textwidth]{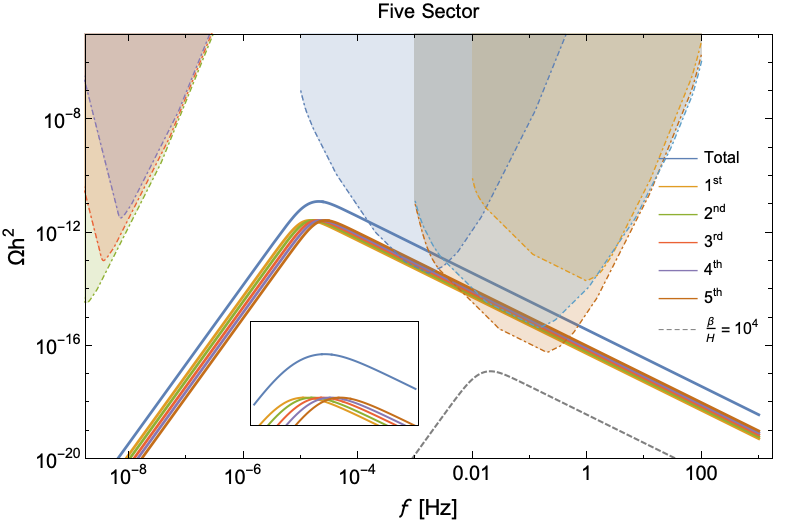} 
\end{minipage}
\hfill
\caption{ Gravitational wave spectral energy density for the various scenarios found in Table \ref{tab:cases}.  All contributions are assumed to be purely from runaway bubble collisions $\Omega_{\phi}$. The colored solid lines use $\beta/H  = 10$ where as the dashed gray line is the total contribution of all sectors for $\beta/H  = 10^4$. The inset is a closer look at the region around the peaks for the $\beta/H  = 10$ case. The shaded curves are the same as Fig.~\ref{fig::Nnatural}.   }
\label{fig:Haa}
\end{figure*}

The GW results of these cases are presented in Fig.~\ref{fig:Haa}. In all cases, the summed GW signal is detectable by one or more proposed interferometers. When changing the assumptions on $\beta/H$, the scenarios in Fig.~\ref{fig:Haa} are still detectable for values ranging between $\mathcal{O}(1)$ and $\mathcal{O}(100)$. As $\beta/H$ increases (decreases) the peak frequency moves to higher (lower) frequencies, dictated by Eq.~(\ref{eqn:peakFrequency}), whereas the amplitude decreases (increases) shown in Eq.~(\ref{eqn::GWengdensity}).

The frequency dependence in Eq.~(\ref{eqn::spectralshape}) takes the form of $f/f_{p}$, this causes a cancellation between the redshifting factors. As multiple sectors phase transition at different times, and therefore different SM photon temperatures, the peaks will shift relative to each other, purely from the linear temperature dependence of the peak frequency $f_{p} \sim T_{\gamma}$ given in Eq.~(\ref{eqn:peakFrequency}).  This is seen in Fig.~\ref{fig:Haa}, where the spectrum peaks are shifted causing a peak broadening of the summed spectrum. The broadening can be substantial if the hidden sectors transition between a large gap of time (temperature). Eventually, a temperature limit will be reached where two (or multiple) distinct peaks will be visible, provided that the amplitudes are comparable. 

\subsection{Detection of Stochastic Graviational Waves}
\label{sec:detection}
A \textit{stochastic} gravitational wave background could be detectable if the signal-to-noise ratio (SNR) is above some threshold value, $\rho > \rho_{th}$, dictated by the capabilities of future interferometers and pulsar timing arrays (PTAs). These interferometers or PTAs quote their experimental sensitivies in terms of spectral noise curves, $S_{\textrm{eff}}(f)$, which can be translated into units of energy density through $h^2 \Omega_{\textrm{eff}}(f) = \frac{2\pi^2}{3H^2}f^3 S_{\textrm{eff}}(f)$. If the experiment uses a single (multiple) detector, the autocorrelated (cross-correlated) SNR is used in comparing to the threshold value $\rho_{th}$. The autocorrelated and cross-correlated SNR are explictly given as~\cite{PhysRevD.59.102001},
\begin{equation}\label{eqn::SNR}
\begin{split}
\rho^2 =  \mathcal{T} \int_{f_{\textrm{min}}}^{f_{\textrm{max}}}\textrm{d}f \bigg( \frac{h^2 \Omega_{\textrm{GW}}(f)}{h^2 \Omega_{\textrm{eff}}(f)} \bigg)^2 \;\;\;\;\;\; \textrm{(autocorrelated)}
\\
\rho^2 = 2 \mathcal{T} \int_{f_{\textrm{min}}}^{f_{\textrm{max}}}\textrm{d}f \bigg( \frac{h^2 \Omega_{\textrm{GW}}(f)}{h^2 \Omega_{\textrm{eff}}(f)} \bigg)^2 \;\;\;\: \textrm{(cross-correlated)},
\end{split}
\end{equation}
where $\mathcal{T}$ is the exposure time of the experiment. The integration covers the entire broadband range of frequencies $(f_{\textrm{min}}, f_{\textrm{max}})$. LISA~\citep{Audley:2017drz} and B-DECIGO~\citep{10.1093/ptep/pty078} are proposed to be single-detector interferometers, whereas BBO~\citep{PhysRevD.72.083005} and DEICIGO~\citep{Sato_2017} would be built from an array of multiple interferometers.
GW signals produced from an early cosmological phase transition would be seen as a stochastic background. Assuming that the GW follows a power law background in frequency, it is commonplace to quote the power law integrated (PLI) sensitivity curves~\cite{PhysRevD.88.124032}. The PLI curves are constructed using information from the power law form of the signal,
\begin{equation}\label{eqn::PowerLaw}
h^2 \Omega_{\textrm{GW}}(f) = h^2 \Omega_{\gamma}  \bigg(\frac{f}{f_{\textrm{ref}}}\bigg)^{\gamma}
\end{equation}
where $\gamma$ is the spectral index of the power law, and $f_{\textrm{ref}}$ is an arbitrary reference frequency which has no effect on the PLI sensitivities.  $h^2 \Omega_{\gamma} $ is the energy density calculated using Eq.~(\ref{eqn::SNR}) with spectral index $\gamma$  and reference frequency $f_{\textrm{ref}}$. The method of calculating the PLI curves involves plotting $h^2 \Omega_{\textrm{GW}}(f) $, using Eq.~(\ref{eqn::PowerLaw}), for various spectral indices $\gamma$ and for some fixed threshold value of $\rho_{th}$. Each curve will lay tangent to the PLI curve, more formally,
\begin{equation}
h^2 \Omega_{\textrm{PLI}} = \max\limits_{\gamma}\bigg[ h^2 \Omega_{\gamma}\bigg(\frac{f}{f_{\textrm{ref}}}\bigg)^{\gamma}  \bigg].
\end{equation}
The spectral noise curves used to create the PLI curves shown in Figs.~\ref{fig::Nnatural} and \ref{fig:Haa} were taken 
from~\citep{Robson_2019,PhysRevD.83.084036,doi:10.1142/S0218271813410137,10.1093/ptep/pty078, Breitbach:2018ddu} for the interferometers and~\citep{Breitbach:2018ddu, Janssen:2014dka} 
for the Square Kilometer Array (SKA) pulsar timing array. 
We have assumed an observation time of $\mathcal{T} = 4 $ years for the interferometers and $\mathcal{T}= 5, 10, 20 $  years for the various stages of SKA. 
In the case of the PTA experiments, the sensitivity curves are dependent on how frequently the pulsar's timing residuals, $\delta t$, are measured. 
When using Eq.~(\ref{eqn::SNR}) to construct the PLI curves for SKA, the upper integration bound is inversely proportional to pulsar's timing residual, $f_{min} = 1/\delta t$.  In this work, it is assumed that $\delta t = 14$ days, but this may underestimate the capabilities of SKA as well as the cadences of the pulsar populations. If the timing residuals are lowered the maximum frequency reach of SKA increases, and the corresponding PLI curves in Figs.~\ref{fig::Nnatural} $\&$ \ref{fig:Haa} are shifted to the right, possibly giving the PTAs sensitivity to some of the scenarios considered here.

\section{Conclusion}
\label{sec:conclusion}

As detection capabilities increase, gravitational wave signals continue to grow in importance as phenomenological signatures that can offer us a unique glimpse into the universe as it was in the early epochs. The space-based interferometers planned for the next generation of GW experiments will be sensitive enough to begin searching for signals of the cataclysmic disruption of space-time due to SFOPT. As we inch closer to these measurements becoming available, it becomes important to develop ways to analyze and understand this data.

Here, we examined scenarios, including $N$naturalness, that involve multiple hidden sectors and calculated the GW profiles present. Our GW projections demonstrate that although $N$naturalness with the reheaton scenario presented in~\cite{Arkani-Hamed:2016rle} is not projected to be detectable in the near future, more generalized scenarios with multiple hidden sector SFOPTs are in an observable region and will begin to be probed by next-generation space experiments. Both cases feature important parts of their GW signals in the void between frequencies detectable by pulsar timing arrays and space-based interferometers --- providing theoretical impetus for new experiments capable of probing this region of frequency space. 

Further, our results provide a framework for understanding and using GW signals in two different ways: first as a unique signal for specific theories featuring multiple SFOPTs and also as a challenge to broaden the understanding of GW detector sensitivity. 

In the former case, this demonstrates the power of GW signals to probe deep into the unknown arena of complex hidden sectors. Individual SFOPTs are understood to create GWs that are assumed to follow an approximate power law. If a model predicts the presence of two, five, or more additional sectors, or features a single extra sector with multiple PTs, deviations from a standard power law can occur. The multiple transitions that occur in the models outlined here create signals that follow this trend: although the individual GWs do obey approximate power laws, their sum does not --- leading to a unique signal indicating so-called dark complexity. Explicitly, a broadening or distortion of the signal around the peak frequency, precisely where the signal has the most energy, could point to a multi-SFOPT scenario and gently guide us in the direction of multiple hidden sectors. 

Shifting to the other part of our framework, our results leads to the question ``how well can experiments probe non-power-law signals?" For frequency ranges away from the peak of the total, GW signals the quoted detection thresholds should hold: the signals fall off as a power law to a very good approximation. However, for areas around the peak frequency the answer is less clear; the PLI curves are built under the assumption of a power law. Work has been done \cite{alanne2019fresh} in examining GW signals using peak amplitudes and peak frequencies as the defining observables: this is rooted in the assumption that GW signals have a model-independent spectral shape around peak frequencies. However, our results indicate that this assumption of model independence cannot hold for all cases: sectors with similar (but different) transition temperatures can create either peak broadening or multihump features that differ significantly from a standard power law shape.  This points to the need for future work to better understand where the power law approximation breaks down and how this affects detection prospects for the next generation of GW detectors. 


\section*{Acknowledgements}

We thank Yang Bai, Djuna Croon, Kevin Earl, and Yuhsin Tsai for helpful discussions. P.A.S. and D.L. both thank the 2018 Theoretical Advanced Study Institute in Elementary Particle Physics Program (TASI) for setting up a platform for discussion which seeded the genesis of this project. This work was supported in part by the Natural Sciences and Engineering Research Council of Canada (NSERC). D.L. acknowledges support from the NSERC Postgraduate Scholarships-Doctoral Program (PGS D).

\appendix
\section*{Appendix: Nonrunaway phase transitions}
\label{sec:appendix}

If the phase transition occurs in the nonrunaway regime $(\alpha_{\infty} > \alpha)$, the dominant contributions to the GW energy densities are given by the sound wave $h^2\Omega_{v}$ and MHD $h^2\Omega_{turb}$ components \citep{Breitbach:2018ddu},
\begin{equation}\label{eqn:Spectralnonrun}
h^2\Omega_{\textrm{GW}} \approx  h^2\Omega_{v} + h^2\Omega_{turb}.
\end{equation}

\begin{figure*}[tb]
\centering
\begin{minipage}[c]{\textwidth}
\includegraphics[width=.45\textwidth ]{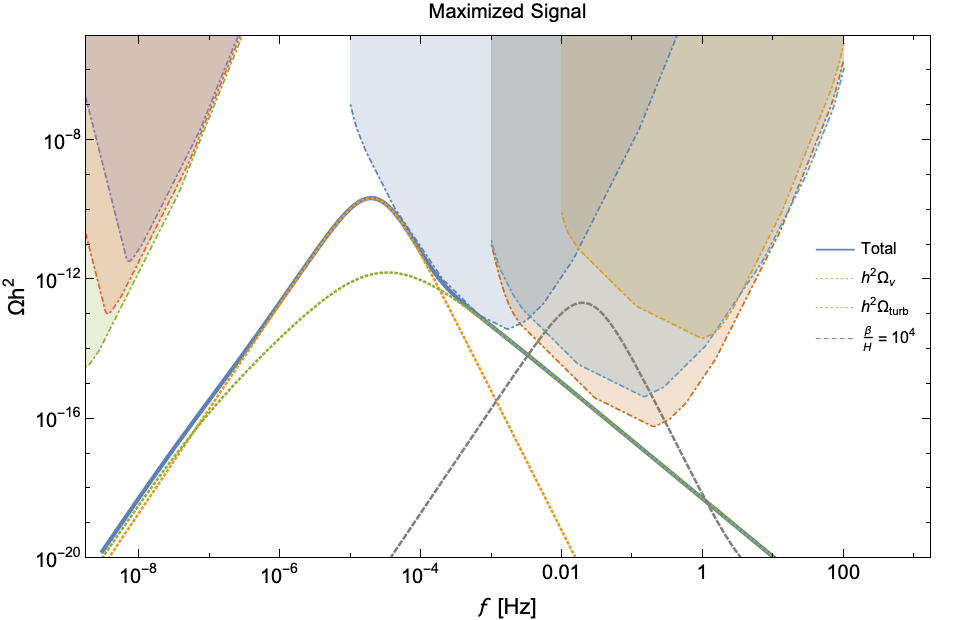}
\hfill
\includegraphics[width=.45\textwidth]{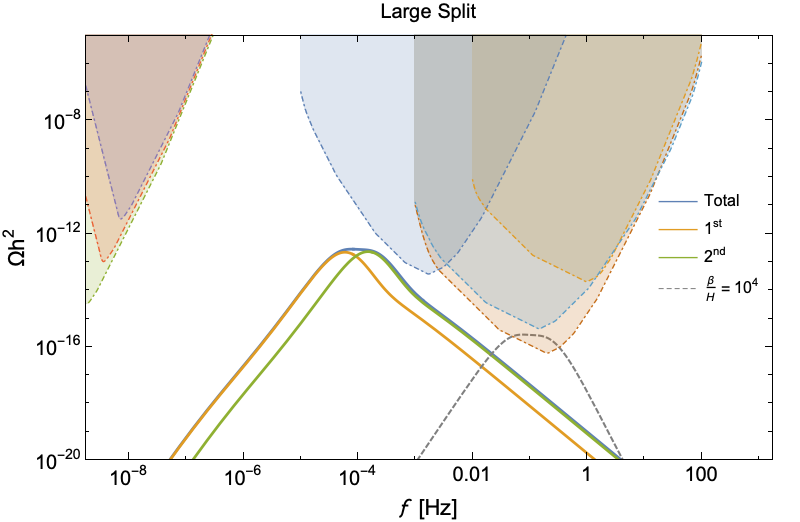} 
\hfill
\includegraphics[width=.45\textwidth]{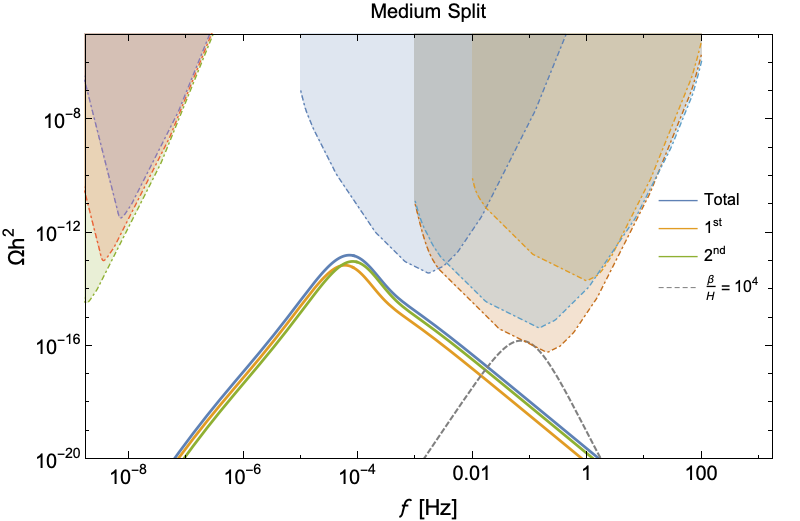} 
\hfill
\includegraphics[width=.45\textwidth]{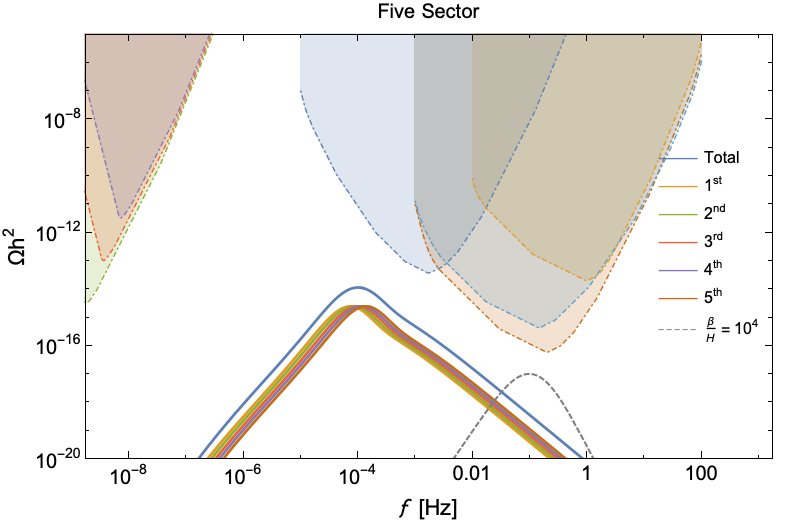} 
\end{minipage}
\hfill
\caption{ Gravitational wave spectral energy density for the various scenarios found in Table \ref{tab:cases}.  All contributions are assumed to be from a nonrunaway phase transition with terminal velocity $v = 0.95$. The coloured solid lines use $\beta/H  = 10$ whereas the dashed grey line is the total contribution of all sectors for $\beta/H  = 10^4$ . The top left figure shows the individual sound wave and MHD contributions.  The shaded curves are the same as Fig.~\ref{fig::Nnatural}. In contrast to the runaway case, most scenarios evade the projected sensitivities. }
\label{fig:Haasw}
\end{figure*}

The new contributions to the GW energy density take on a different form from Eq.~(\ref{eqn::GWengdensity}) \citep{Caprini:2015zlo},

\begin{equation}\label{eqn::SWMHD}
\begin{split}
h^2\Omega_{v}^* = 1.6 \times 10^{-1} \, v  \, \bigg( \frac{\kappa_{v}\;  \alpha}{1 + \alpha} \bigg)^2 \bigg( \frac{H}{\beta} \bigg)^1  S_{v}(f),
\\
h^2\Omega_{turb}^* = 2.01\times 10^{1} \, v \, \bigg( \frac{\kappa_{turb} \; \alpha}{1 + \alpha} \bigg)^{3/2} \bigg( \frac{H}{\beta} \bigg)^1  S_{turb}(f).
\end{split}
\end{equation}
Unlike the runaway case, we do not assume $v = 1$ due to the bubbles reaching a terminal velocity. The MHD efficiency factor is a fraction of the sound waves, $\kappa_{turb} = \epsilon\kappa_{v}$. Current simulations have motivated a range of $\epsilon \sim 0.05 - 0.10$  \citep{Caprini:2015zlo}, where we take the optimistic case of $\epsilon = 0.10$. Both contributions have unique spectral shapes, given to be \citep{Breitbach:2018ddu}, 
\begin{equation}
\begin{split}
S_{v}(f) = (f/f_{p,v})^3\bigg(\frac{7}{4+3(f/f_{p,v})^2}\bigg)^{7/2},
\\
S_{turb}(f) = \frac{(f/f_{p,turb})^{3}}{(1 + f/f_{p,turb})^{11/3} (1 + 8\pi(f/H))}.
\end{split}
\end{equation}
The MHD energy density has a spectral shape dependent on the Hubble rate at the time of nucleation, $H$.  Similar to the scalar spectral shape in Eq.~(\ref{eqn::spectralshape}), the frequencies are scaled by their respective temperature-dependent peak frequency: 

\begin{equation}
\begin{split}
f_{p,v} = 1.9 \times 10^{-5} \, \textrm{Hz}\, \frac{1}{v} \bigg( \frac{\beta}{H}\bigg)\bigg(\frac{T_{\gamma}}{100 \, \textrm{GeV}}\bigg)\bigg(\frac{g_{*}}{100}\bigg)^{\frac{1}{6}},
\\
f_{p,turb} =  2.7 \times 10^{-5} \, \textrm{Hz}\, \frac{1}{v} \bigg( \frac{\beta}{H}\bigg)\bigg(\frac{T_{\gamma}}{100 \, \textrm{GeV}}\bigg)\bigg(\frac{g_{*}}{100}\bigg)^{\frac{1}{6}}.
\end{split}
\end{equation}
We evolve the frequencies and energy densities with the same redshift factors, Eq.~(\ref{eqn::redshifting}) from Sec~\ref{sec:gw}. Fig.~\ref{fig:Haasw} shows the sum of the GW energy density for the scenarios in Table~\ref{tab:cases}, but instead for the nonrunaway case, with a terminal velocity of $v = 0.95$. For this case we see that the spectra tend to be shifted to lower frequencies and are more likely to fall in the gap between the interferometers and the pulsar-based detectors. On the other hand, larger values of $\beta/H$ increase the typical frequency, so this case becomes more sensitive in some scenarios to values of $\beta/H$ on the larger end of the considered range.

In the sound wave case, these parameterizations are extracted from simulations with $\beta/H < 100$ corresponding to a long-lasting sound wave component. For high $\beta/H$ the transition timescale from sound wave to MHD turbulence is much shorter than the Hubble time. When estimating the model expectations for $\beta/H = 10^4$, we enter a regime at which Eq.~(\ref{eqn::SWMHD}) may be overestimating the sound wave contribution. Investigations of this regime have been done in \citep{Hindmarsh:2017gnf, Ellis:2018mja, Ellis_2019}. This effect, however, only affects the amplitude of the signal, and our work focuses on the unique spectral shapes that are formed in these models. Therefore we project our results for high $\beta/H$ in Fig.~\ref{fig:Haasw} to motivate the novel spectral profiles.

Finally, we note that because the sound wave and MHD contributions have different spectral shapes, the overall spectrum has a kink at a frequency above the peak. In the top left panel of Fig.~\ref{fig:Haasw}, we show the two contributions separately in addition to their sum to highlight this effect.

\appendix
\bibliographystyle{apsrev4-1}
\bibliography{refs}
\end{document}